\documentstyle[aps,epsfig]{revtex}

\begin{document}
\title{Proton spin structure and valence quarks\protect\footnote
{to be published in Phys. Rev. D}}
\author{Petr Z\'avada}
\address{Institute of Physics, Academy of Sciences of the Czech Republic,\\
Na Slovance 2, CZ-182 21 Prague 8, \\
e-mail: zavada@fzu.cz \\
}
\date{November 18, 2002}
\maketitle
\pacs{13.60.-r, 13.88.+e, 14.65.-q}

\begin{abstract}
The spin structure of the system of quasifree fermions having total angular
momentum $J=1/2$ is studied in a consistently covariant approach. Within
this model the relations between the spin functions are obtained. Their
particular cases are the sum rules Wanzura - Wilczek, Efremov - Leader -
Teryaev, Burkhardt - Cottingham and also the expression for the Wanzura -
Wilczek twist 2 term $g_{2}^{WW}$. With the use of the proton valence quark
distributions as an input, the corresponding spin functions are obtained.
The resulting structure functions $g_{1}$ and $g_{2}$ are well compatible
with the experimental data.
\end{abstract}

\section{Introduction}

\label{sec1}The nucleon structure functions, both unpolarized and polarized,
are the basic tools for understanding the nucleon internal structure in the
language of the QCD. Precision measurements on the polarized structure
functions have been completed only in the recent years \cite{e142} -\cite%
{her2}. These functions contain information which is a crucial complement to
the structure functions obtained in the unpolarized deep inelastic
scattering (DIS) experiments. At the same time the interpretation and
understanding of polarized structure functions seem to be much more
difficult than in the case of unpolarized ones. Actually, until now it is
not clear how the nucleon spin is generated from the spins and orbital
momenta of the quarks and gluons. For the present status and perspectives of
the nucleon spin physics see \cite{stoss} and citations therein. The more
formal aspects of the polarized DIS are explained in \cite{efrem1} and \cite%
{barone}.

The spin in general is a very delicate quantity, which requires
correspondingly precise treatment. It has been argued, that for correct
evaluation of the quark contribution to the nucleon spin it is necessary to
take properly into account the intrinsic quark motion \cite{bro} - \cite%
{zav4}. The necessity of the covariant formulation of the quark - parton
model (QPM) for the spin functions has been pointed out in \cite{cfs}. These
requirements are not satisfied in the standard formulation of the QPM, which
is currently used for analysis and interpretation of the experimental data.

In the paper \cite{zav4} we have demonstrated the role of the intrinsic
motion for the spin structure functions, using a simple model of the system
quasifree fermions on mass shell. The basic requirement was a consistently
covariant formulation of the task for the system of fermions, which are not
static, being characterized by some momenta distribution in the frame of
their center of mass. In the present paper we attempt to further develop
this approach. In section \ref{sec2} we are introducing the spin structure
functions $g_{1}(x),g_{2}(x)$, the longitudinal and transversal net spin
density distributions $s_{L}(x),s_{T}(x)$, and the density of total angular
momentum $j(x)$. Then in Sec. \ref{sec3} we show how all of these functions
are mutually related. Finally in Sec. \ref{sec4} \ we apply the suggested
approach, with some simplified assumptions, to the description of the proton
spin structure and make a comparison with the experimental data on the $%
g_{1}(x)$ and $g_{2}(x)$. The last section is devoted to the short summary
and conclusion.

\section{Spin structure functions and spin distributions}

\label{sec2}In the previous paper \cite{zav4} we have shown that the spin
structure functions, related to the spherically symmetric target consisting
of the three quasifree fermions (of spin $1/2$), having resulting total
angular momentum $J=1/2$, can be written as:%
\begin{equation}
g_{1}=\frac{1}{2}\int H(p_{0})\left[ m+\frac{\nu }{\left| {\bf q}\right| }%
p_{1}+\frac{\nu ^{2}}{\left| {\bf q}\right| ^{2}}\frac{p_{1}^{2}}{p_{0}+m}%
+\left( 1-\frac{\nu ^{2}}{\left| {\bf q}\right| ^{2}}\right) \frac{%
p_{T}^{2}/2}{p_{0}+m}\right] \delta \left( \frac{p_{0}\nu +p_{1}\left| {\bf q%
}\right| }{M\nu }-x\right) \frac{d^{3}p}{p_{0}},  \label{t2}
\end{equation}%
\begin{equation}
g_{2}=-\frac{1}{2}\frac{\nu }{\left| {\bf q}\right| }\int H(p_{0})\left(
p_{1}+\frac{\nu }{\left| {\bf q}\right| }\frac{p_{1}^{2}-p_{T}^{2}/2}{p_{0}+m%
}\right) \delta \left( \frac{p_{0}\nu +p_{1}\left| {\bf q}\right| }{M\nu }%
-x\right) \frac{d^{3}p}{p_{0}},  \label{t3}
\end{equation}%
where $p_{0}=\sqrt{m^{2}+{\bf p}^{2}}$, $p_{T}^{2}=p_{2}^{2}+p_{3}^{2}$, and 
$H$ is the charge weighted distribution%
\begin{equation}
H(p_{0})=\sum_{k=1}^{3}e_{k}^{2}\Delta G_{k}(p_{0}).  \label{t4}
\end{equation}%
This distribution is constructed from the polarized distributions of
individual fermions%
\begin{equation}
\Delta G_{k}(p_{0})\equiv G_{k,+1/2}(p_{0})-G_{k,-1/2}(p_{0}),  \label{t5}
\end{equation}%
which satisfy%
\begin{equation}
\int G_{k}(p_{0})d^{3}p=1;\qquad G_{k}(p_{0})\equiv
G_{k,+1/2}(p_{0})+G_{k,-1/2}(p_{0}),  \label{ta5}
\end{equation}%
\begin{equation}
\int \Delta G(p_{0})d^{3}p=1;\qquad \Delta G(p_{0})=\sum_{k=1}^{3}\Delta
G_{k}(p_{0}).  \label{tb5}
\end{equation}%
Distributions $G_{k,\lambda }(p_{0})$ measure probability to find a fermion
in the state%
\begin{equation}
u\left( p,\lambda {\bf n}\right) =\frac{1}{\sqrt{N}}\left( 
\begin{array}{c}
\phi _{\lambda {\bf n}} \\ 
\frac{{\bf p}{\bf \sigma }}{p_{0}+m}\phi _{\lambda {\bf n}}%
\end{array}%
\right) ;\qquad \frac{1}{2}{\bf n\sigma }\phi _{\lambda {\bf n}}=\lambda
\phi _{\lambda {\bf n}},\qquad \lambda =\pm \frac{1}{2},  \label{t6}
\end{equation}%
where the direction ${\bf n}$ coincides with the direction of target
polarization ${\bf J}$ and a standard normalization is used:%
\begin{equation}
N=\frac{2p_{0}}{p_{0}+m},\qquad \phi _{\lambda }^{\dagger }\phi _{\lambda
}=1.  \label{t8}
\end{equation}

Now, let us try using the same system to calculate some spin distribution
functions. In the first step we shall define these distributions in terms of
the fermion momenta related to the target rest frame, then we shall show
their representation in the variable $x$.

The net spin density corresponding to the projection on the direction ${\bf n%
}^{\prime }$ is defined as%
\begin{equation}
S({\bf p,n,n}^{\prime })=\sum_{\lambda }G_{\lambda }(p_{0})u^{\dagger
}(p,\lambda {\bf n})\left( {\bf n}^{\prime }{\bf \Sigma }\right) u(p,\lambda 
{\bf n});\qquad {\bf \Sigma }=\frac{1}{2}\left( 
\begin{array}{cc}
{\bf \sigma } & {\bf 0} \\ 
{\bf 0} & {\bf \sigma }%
\end{array}%
\right) .  \label{t19}
\end{equation}%
One can verify, that this expression can be modified%
\begin{equation}
S({\bf p,n,n}^{\prime })=\frac{1}{N}\sum_{\lambda }G_{\lambda }(p_{0})\left(
\lambda {\bf nn}^{\prime }-\lambda {\bf nn}^{\prime }\frac{{\bf p}^{2}}{%
\left( p_{0}+m\right) ^{2}}+\phi _{\lambda {\bf n}}^{\dagger }\frac{{\bf p}%
{\bf \sigma }\cdot {\bf pn}^{\prime }}{\left( p_{0}+m\right) ^{2}}\phi
_{\lambda {\bf n}}\right) .  \label{tx19}
\end{equation}%
We assume, that the beam direction is defined by the vector ${\bf k}=(\left| 
{\bf k}\right| ,0,0)$, then one can obtain the following particular cases of
the distribution (\ref{t19}).

1) Longitudinal polarization in longitudinally polarized target, i.e. ${\bf n%
}={\bf n}^{\prime }=(1,0,0)$, then%
\begin{equation}
{\bf p}{\bf \sigma }\cdot {\bf pn}^{\prime }=\sum_{i=1}^{3}p_{i}^{2}\sigma
_{i}n_{i}^{\prime }+\sum_{j\neq i}p_{i}p_{j}\sigma _{i}n_{j}^{\prime
}=p_{1}^{2}\sigma _{1}+p_{2}p_{1}\sigma _{2}+p_{3}p_{1}\sigma _{3}
\label{ty19}
\end{equation}%
and the relation (\ref{tx19}) can be simplified%
\begin{equation}
S_{L}({\bf p})=\frac{1}{2p_{0}}\Delta G(p_{0})\left( m+\frac{p_{1}^{2}}{%
p_{0}+m}\right) .  \label{t22}
\end{equation}

2) Transverse polarization in transversely polarized target, i.e. ${\bf n}=%
{\bf n}^{\prime }=(0,1,0)$, then 
\begin{equation}
{\bf p}{\bf \sigma }\cdot {\bf pn}=\sum_{i=1}^{3}p_{i}^{2}\sigma
_{i}n_{i}+\sum_{j\neq i}p_{i}p_{j}\sigma _{i}n_{j}=p_{2}^{2}\sigma
_{2}+p_{1}p_{2}\sigma _{1}+p_{3}p_{2}\sigma _{3}  \label{t20}
\end{equation}%
and the relation (\ref{tx19}) can be simplified%
\begin{equation}
S_{T}({\bf p})=\frac{1}{2p_{0}}\Delta G(p_{0})\left( m+\frac{p_{2}^{2}}{%
p_{0}+m}\right) .  \label{t21}
\end{equation}

3) In a similar way, one can also obtain the polarizations $S_{L}^{T}({\bf p}%
)$ and $S_{T}^{L}({\bf p})$, which are related to the density of
longitudinal polarization in the transversely polarized target and vice
versa. The density $S_{L}^{T}({\bf p})$ can be obtained from the relation (%
\ref{tx19}) after inserting ${\bf n=}(0,1,0),\ {\bf n}^{\prime }{\bf =}%
(1,0,0)$ and $S_{T}^{L}({\bf p})$ with ${\bf n=}(1,0,0),\ {\bf n}^{\prime }%
{\bf =}(0,1,0)$ correspondingly. After some calculation similar to that for
obtaining the relations (\ref{t22}) and (\ref{t21}), one gets%
\begin{equation}
S_{L}^{T}({\bf p})=S_{T}^{L}({\bf p})=\frac{1}{2p_{0}}\Delta G(p_{0})\frac{%
p_{1}p_{2}}{p_{0}+m}.  \label{t28}
\end{equation}

4) The density of total angular momentum can be defined as:%
\begin{equation}
J({\bf p,n,n}^{\prime })=\sum_{\lambda }G_{\lambda }(p_{0})u^{\dagger
}(p,\lambda {\bf n})\left( {\bf n}^{\prime }{\bf j}\right) u(p,\lambda {\bf n%
});\qquad j_{k}=\Sigma _{k}+l_{k}=\frac{1}{2}\left( 
\begin{array}{cc}
\sigma _{k} & 0 \\ 
0 & \sigma _{k}%
\end{array}%
\right) -i\varepsilon _{klm}p_{l}\frac{\partial }{\partial p_{m}}.
\label{tx28}
\end{equation}%
One can verify, that after some calculation this expression can be
simplified:%
\begin{equation}
J({\bf p,n,n}^{\prime })=\frac{1}{2}{\bf nn}^{\prime }\Delta G(p_{0}).
\label{ty28}
\end{equation}%
This result implies, that $J$ has rotational symmetry, so there is no
distinction between longitudinal and transversal density:%
\begin{equation}
J_{L}({\bf p})=J_{T}({\bf p})=\frac{1}{2}\Delta G(p_{0})\equiv J({\bf p}%
),\qquad J_{L}^{T}({\bf p})=J_{T}^{T}({\bf p})=0.  \label{tz28}
\end{equation}%
\ 

Further, is it possible to express the obtained distributions as the
functions of $x$ instead of $p$? For a simplification we shall from now on
assume that%
\begin{equation}
Q^{2}\gg 4M^{2}x^{2},  \label{t9}
\end{equation}%
which implies%
\begin{equation}
\frac{\left| {\bf q}\right| }{\nu }=\sqrt{1+4M^{2}x^{2}/Q^{2}}\rightarrow 1.
\label{t10}
\end{equation}%
Then the $\delta $ function term, which defines the transformation $%
p\rightarrow x$ in Eqs. (\ref{t2}), (\ref{t3}) will be simplified%
\begin{equation}
\delta \left( \frac{p_{0}\nu +p_{1}\left| {\bf q}\right| }{M\nu }-x\right)
\rightarrow \delta \left( \frac{p_{0}+p_{1}}{M}-x\right) ,  \label{t11}
\end{equation}%
in this limit the coordinate $p_{1}$ defines the beam direction. The spin
structure functions (\ref{t2}), (\ref{t3}) are now simplified accordingly:%
\begin{equation}
g_{1}(x)=\frac{1}{2}\int H(p_{0})\left( m+p_{1}+\frac{p_{1}^{2}}{p_{0}+m}%
\right) \delta \left( \frac{p_{0}+p_{1}}{M}-x\right) \frac{d^{3}p}{p_{0}},
\label{t13}
\end{equation}%
\begin{equation}
g_{2}(x)=-\frac{1}{2}\int H(p_{0})\left( p_{1}+\frac{p_{1}^{2}-p_{T}^{2}/2}{%
p_{0}+m}\right) \delta \left( \frac{p_{0}+p_{1}}{M}-x\right) \frac{d^{3}p}{%
p_{0}},  \label{t14}
\end{equation}%
\begin{equation}
g_{1}(x)+g_{2}(x)=\frac{1}{2}\int H(p_{0})\left( m+\frac{p_{T}^{2}/2}{p_{0}+m%
}\right) \delta \left( \frac{p_{0}+p_{1}}{M}-x\right) \frac{d^{3}p}{p_{0}}.
\label{ta14}
\end{equation}%
Apparently, the convolution defined by the $\delta $ function (\ref{t11})
also gives the rule for transformation of the spin distributions expressed
in the variable $p$ to the corresponding representation in the variable $x$:%
\begin{equation}
j(x)=\int J({\bf p})\delta \left( \frac{p_{0}+p_{1}}{M}-x\right) d^{3}p=%
\frac{1}{2}\int \Delta G(p_{0})\delta \left( \frac{p_{0}+p_{1}}{M}-x\right)
d^{3}p,  \label{t12}
\end{equation}%
\begin{equation}
s_{L}(x)=\int S_{L}({\bf p})\delta \left( \frac{p_{0}+p_{1}}{M}-x\right)
d^{3}p=\frac{1}{2}\int \Delta G(p_{0})\left( m+\frac{p_{1}^{2}}{p_{0}+m}%
\right) \delta \left( \frac{p_{0}+p_{1}}{M}-x\right) \frac{d^{3}p}{p_{0}},
\label{t24}
\end{equation}%
\begin{equation}
s_{T}(x)=\int S_{T}({\bf p})\delta \left( \frac{p_{0}+p_{1}}{M}-x\right)
d^{3}p=\frac{1}{2}\int \Delta G(p_{0})\left( m+\frac{p_{T}^{2}}{2\left(
p_{0}+m\right) }\right) \delta \left( \frac{p_{0}+p_{1}}{M}-x\right) \frac{%
d^{3}p}{p_{0}}.  \label{t23}
\end{equation}%
In the last integral we could replace $p_{2}^{2}$ by $p_{T}^{2}/2$ because
of the axial symmetry. Further, it is obvious, that the densities $S_{L}^{T}$
and $S_{T}^{L}$, expressed in the variable $x$, vanish due to the symmetry:%
\begin{equation}
s_{L}^{T}(x)=s_{T}^{L}(x)=\frac{1}{2}\int \Delta G(p_{0})\frac{p_{1}p_{2}}{%
p_{0}+m}\delta \left( \frac{p_{0}+p_{1}}{M}-x\right) \frac{d^{3}p}{p_{0}}=0.
\label{t29}
\end{equation}

What is the meaning of the integrals in the relations (\ref{t13}) - (\ref%
{t23})? To simplify this question, let us assume the same shape of the
distributions $G_{k}(p_{0})$ for all the three fermions. Then the
distributions $\Delta G$ and $H$ differ only by a constant factor, in which
charges and polarizations of individual fermions are absorbed:%
\begin{equation}
H(p_{0})=\kappa \Delta G(p_{0});\qquad \kappa =\sum_{k=1}^{3}e_{k}^{2}\int
\Delta G_{k}(p_{0})d^{3}p.  \label{t15}
\end{equation}

Now, in agreement with the results obtained in \cite{zav4}, one can observe
the following. The relation (\ref{t13}) can be rewritten%
\begin{equation}
g_{1}(x)=\frac{\kappa }{2}\int \Delta G(p_{0})\left( m+p_{1}+\frac{p_{1}^{2}%
}{p_{0}+m}\right) \delta \left( \frac{p_{0}+p_{1}}{M}-x\right) \frac{d^{3}p}{%
p_{0}}  \label{t16}
\end{equation}%
and after integration over $x$ one gets%
\begin{equation}
\Gamma _{1}=\int g_{1}(x)dx=\frac{\kappa }{2}\int \Delta G(p_{0})\left( 
\frac{1}{3}+\frac{2m}{3p_{0}}\right) d^{3}p=\kappa \left\langle {\bf n\Sigma 
}\right\rangle ,  \label{t17}
\end{equation}%
where ${\bf n}\ $is the direction of the target polarization and $%
\left\langle {\bf n\Sigma }\right\rangle $ represents the resulting
projection of the spins coming from the individual fermions. In a similar
way one gets: 
\begin{equation}
\int j(x)dx=\frac{1}{2}\int \Delta G(p_{0})d^{3}p=\frac{1}{2},  \label{t18}
\end{equation}%
i.e. the integral represents the resulting projection of the total angular
momentum. Further, one can easily check: 
\begin{equation}
\int s_{L}(x)dx=\int s_{T}(x)dx=\frac{1}{2}\int \Delta G(p_{0})\left( \frac{1%
}{3}+\frac{2m}{3p_{0}}\right) d^{3}p=\frac{\Gamma _{1}}{\kappa }.
\label{t25}
\end{equation}%
Moreover, the following relation is valid:%
\begin{equation}
\kappa \cdot s_{T}(x)=g_{1}(x)+g_{2}(x).  \label{t26}
\end{equation}

Let us note, despite our assumption that the target consists of just the
three fermions, the suggested approach is more general. Since the spin
functions are always based on the differences like (\ref{t5}), all the
resulting relations are equally valid for any target consisting of the
fermions with $\Delta G_{k}(p_{0})\neq 0;\quad k=1,2,3$, which are embedded
in another system with compensated spins: $\Delta G_{k}(p_{0})=0;\quad
k=4,5,6,...$

Now we can summarize:

{\it i)} The function $j(x)$ measures the contribution of total angular
momenta (spin + orbital momentum) of the constituent fermions to the target
spin.

{\it ii)} The functions $s_{L(T)}(x)$ measure the net spin contribution of
the constituent fermions to the spin of target with longitudinal
(transversal) polarization. Apparently, one can calculate also the
corresponding densities of the orbital momentum as%
\begin{equation}
l_{L(T)}=j(x)-s_{L(T)}(x).  \label{ta26}
\end{equation}

{\it iii) }Obviously the functions $g_{1}(x),j(x),s_{L}(x)$ and $s_{T}(x)$
are equivalent in the case of the static fermions, where orbital momentum
does not play any role. The considered distribution functions $\Delta G$ and 
$H$ have rotational symmetry in the target rest frame, which is a necessary
condition for the target with spin $J=1/2$. It follows, that meaning of our
function $j(x)$ suggested above {\it does not depend} on the orientation of
the target polarization (longitudinal or transversal) with respect to the
beam direction. Let us point out, the last statement can be deduced only in
the framework of the relativistically covariant description, in which the
rotational symmetry of the target is properly taken into account. At the
same time, let us note that in general, the function $g_{1}(x)$ is not
equivalent to the measure of the longitudinal spin density $s_{L}(x)$, only
their integrals over $x$ are equal (up to factor $\kappa $).

\section{Relationship among the spin functions}

\label{sec3}Before next discussion we shall first prove the following
proposition:

The functions $V_{n}(x)$ defined as%
\begin{equation}
V_{n}(x)=\int H(p_{0})\left( \frac{p_{0}}{M}\right) ^{n}\delta \left( \frac{%
p_{0}+p_{1}}{M}-x\right) d^{3}p;\qquad p_{0}=\sqrt{m^{2}+{\bf p}^{2}}
\label{t30}
\end{equation}%
satisfy%
\begin{equation}
\frac{V_{j}^{\prime }(x)}{V_{k}^{\prime }(x)}=\left( \frac{x}{2}+\frac{%
x_{0}^{2}}{2x}\right) ^{j-k};\qquad x_{0}=\frac{m}{M}  \label{T31}
\end{equation}%
for any powers $j,k$ and function $H$, for which the integral (\ref{t30})
exists. Proof of the last relation is given in the Appendix \ref{app1}.

\subsection{Spin structure functions $g_{1}(x),g_{2}(x)$}

{\it \ }\ With the use of the relations (\ref{t30}),(\ref{T31}), as shown in
the Appendix \ref{app2}, one can rewrite the Eqs. (\ref{t14}) and (\ref{ta14}%
) as%
\begin{equation}
g_{2}(x)=-\frac{1}{2}\left[ a(x)V_{0}(x)+\int_{x}^{1}b(x,y)V_{0}(y)dy\right] 
\label{T32}
\end{equation}%
\begin{equation}
g_{1}(x)+g_{2}(x)=\frac{1}{2}\left[ A(x)V_{0}(x)+\int_{x}^{1}B(x,y)V_{0}(y)dy%
\right]   \label{T33}
\end{equation}%
where%
\begin{equation}
a(x)=2x\frac{x-x_{0}}{x^{2}+x_{0}^{2}},\qquad b(x,y)=\left[ \frac{3\left(
x+x_{0}\right) ^{2}}{\left( y+x_{0}\right) ^{4}}-\frac{%
3x^{2}+2xx_{0}+x_{0}^{2}}{\left( y^{2}+x_{0}^{2}\right) ^{2}}\right] \frac{%
y^{2}-x_{0}^{2}}{x_{0}},  \label{t34}
\end{equation}%
\begin{equation}
A(x)=\frac{2xx_{0}}{x^{2}+x_{0}^{2}},\qquad B(x,y)=\left[ -\frac{\left(
x+x_{0}\right) ^{2}}{\left( y+x_{0}\right) ^{4}}+\frac{x^{2}-x_{0}^{2}}{%
\left( y^{2}+x_{0}^{2}\right) ^{2}}\right] \frac{y^{2}-x_{0}^{2}}{x_{0}}.
\label{t35}
\end{equation}%
Now, one can easily check, that in the limit $x_{0}\rightarrow 0$ the
relations (\ref{T32}), (\ref{T33}) are simplified:%
\begin{equation}
g_{2}(x)=-V_{0}(x)+\int_{x}^{1}\left( 6\frac{x^{2}}{y^{3}}-2\frac{x}{y^{2}}%
\right) V_{0}(y)dy,  \label{t36}
\end{equation}%
\begin{equation}
g_{1}(x)+g_{2}(x)=\int_{x}^{1}\left( 2\frac{x^{2}}{y^{3}}-\frac{x}{y^{2}}%
\right) V_{0}(y)dy.  \label{t37}
\end{equation}%
These relations imply%
\begin{equation}
g_{1}(x)=V_{0}(x)-\int_{x}^{1}\left( 4\frac{x^{2}}{y^{3}}-\frac{x}{y^{2}}%
\right) V_{0}(y)dy  \label{t38}
\end{equation}%
and%
\begin{equation}
\left[ g_{1}(x)+g_{2}(x)\right] ^{\prime }=\int_{x}^{1}\left( 4\frac{x}{y^{3}%
}-\frac{1}{y^{2}}\right) V_{0}(y)dy-\frac{V_{0}(x)}{x}.  \label{t39}
\end{equation}%
The combining of the last two relations gives%
\begin{equation}
\left[ g_{1}(x)+g_{2}(x)\right] ^{\prime }=-\frac{g_{1}(x)}{x}  \label{t41}
\end{equation}%
or%
\begin{equation}
g_{2}(x)=-g_{1}(x)+\int_{x}^{1}\frac{g_{1}(y)}{y}dy,  \label{t42}
\end{equation}%
which is the known expression for Wanzura - Wilczek twist-2 term for $g_{2}$
approximation\cite{wawi}.

Can we now obtain a similar relation for the case $x_{0}>0$, i.e. for the
massive fermions? \ Let us combine Eqs. (\ref{T32}), (\ref{T33}) to the form%
\begin{equation}
g_{1}(x)+\left( 1+\frac{A(x)}{a(x)}\right) g_{2}(x)=\frac{1}{2}%
\int_{x}^{1}\left( B(x,y)-\frac{A(x)}{a(x)}b\left( x,y\right) \right)
V_{0}(y)dy  \label{t43}
\end{equation}%
and let us try to express differentiation of r.h.s. as a combination of $%
g_{1}$ and $g_{2}$:%
\begin{equation}
\left[ \frac{1}{2}\int_{x}^{1}\left( B(x,y)-\frac{A(x)}{a(x)}b\left(
x,y\right) \right) V_{0}(y)dy\right] ^{\prime
}=c_{1}(x)g_{1}(x)+c_{2}(x)g_{2}(x).  \label{t44}
\end{equation}%
In the Appendix \ref{app3} it is shown, that after inserting $g_{1},g_{2}$
from Eqs. (\ref{T32}), (\ref{T33}) this equation is solvable for $%
c_{1}(x),c_{2}(x)$, then after comparing with Eq. (\ref{t43}) we get%
\begin{equation}
\left[ g_{1}(x)+\left( 1+\frac{A(x)}{a(x)}\right) g_{2}(x)\right] ^{\prime
}=c_{1}(x)g_{1}(x)+c_{2}(x)g_{2}(x),  \label{T45}
\end{equation}%
where%
\begin{equation}
c_{1}(x)=-\frac{x^{2}+4xx_{0}+x_{0}^{2}}{\left( x^{2}-x_{0}^{2}\right)
\left( x+2x_{0}\right) },\qquad c_{2}(x)=-\frac{x_{0}\left(
x^{2}+xx_{0}+4x_{0}^{2}\right) }{\left( x-x_{0}\right) \left(
x^{2}-x_{0}^{2}\right) \left( x+2x_{0}\right) }.  \label{t46}
\end{equation}%
In the same Appendix it is shown, that Eq. (\ref{T45}) implies%
\begin{equation}
g_{2}(x)=-\frac{x-x_{0}}{x}g_{1}(x)+\frac{x\left( x+2x_{0}\right) }{\left(
x+x_{0}\right) ^{2}}\int_{x}^{1}\frac{y^{2}-x_{0}^{2}}{y^{3}}g_{1}(y)dy,
\label{t47}
\end{equation}%
\begin{equation}
g_{1}(x)=-\frac{x}{x-x_{0}}g_{2}(x)-\frac{x+2x_{0}}{x^{2}-x_{0}^{2}}%
\int_{x}^{1}g_{2}(y)dy.  \label{T48}
\end{equation}%
One can check, that for $x_{0}\rightarrow 0$ both the last relations are
equivalent to Eq. (\ref{t41}).

Further, in the limit $x_{0}\rightarrow 0$ one can also easily calculate the
momenta of the spin structure functions $g_{1},g_{2}$. If we define 
\begin{equation}
\left\langle x^{\alpha }\right\rangle =\int_{0}^{1}x^{\alpha }V_{0}(x)dx,
\label{t49}
\end{equation}%
then after integrating by parts the following relation is obtained:%
\begin{equation}
\int_{0}^{1}x^{\alpha }\int_{x}^{1}\frac{V_{0}(y)}{y^{\beta }}dy=\frac{%
\left\langle x^{\alpha -\beta +1}\right\rangle }{\alpha +1}.  \label{t50}
\end{equation}%
Application of this relation in Eqs. (\ref{t37}), (\ref{t36}) then gives%
\begin{equation}
\int_{0}^{1}x^{\alpha }\left[ g_{1}(x)+g_{2}(x)\right] dx=\left\langle
x^{\alpha }\right\rangle \frac{\alpha +1}{\left( \alpha +2\right) \left(
\alpha +3\right) },  \label{t51}
\end{equation}%
\begin{equation}
\int_{0}^{1}x^{\alpha }g_{2}(x)dx=-\left\langle x^{\alpha }\right\rangle 
\frac{\alpha \left( \alpha +1\right) }{\left( \alpha +2\right) \left( \alpha
+3\right) }  \label{t52}
\end{equation}%
for {\it any} $\alpha $, for which the integrals exist. The last two
relations imply%
\begin{equation}
\int_{0}^{1}x^{\alpha }\left[ \frac{\alpha }{\alpha +1}g_{1}(x)+g_{2}(x)%
\right] dx=0,  \label{t53}
\end{equation}%
which for $\alpha =2,4,6,...$ corresponds to the Wanzura - Wilczek sum rules %
\cite{wawi}. Other special cases correspond to the Burkhardt - Cottingham ($%
\alpha =0$) and the Efremov - Leader - Teryaev (ELT, $\alpha =1$) sum rules %
\cite{buco}, \cite{efrem2}.

\subsection{Spin distributions $j(x),s_{L}(x)$ and $s_{T}(x)$}

Now, let us try to find the relations among the spin functions $%
j(x),s_{L}(x),s_{T}(x)$ and the structure functions $g_{1},g_{2}$. For this
purpose we only slightly change the definitions (\ref{t12}),(\ref{t24}) (\ref%
{t23}) in which we replace the function $\Delta G(p_{0})$ [Eq. (\ref{tb5}),
the sum of spin contributions from individual fermions] by the function $%
H(p_{0})$ [Eq. (\ref{t4}), spin contributions are weighted by the charges,
see also the paragraph involving Eq. (\ref{t15})]. The new relations read:%
\begin{equation}
j(x)=\frac{1}{2}\int H(p_{0})\delta \left( \frac{p_{0}+p_{1}}{M}-x\right)
d^{3}p,  \label{t54}
\end{equation}%
\begin{equation}
s_{L}(x)=\frac{1}{2}\int H(p_{0})\left( m+\frac{p_{1}^{2}}{p_{0}+m}\right)
\delta \left( \frac{p_{0}+p_{1}}{M}-x\right) \frac{d^{3}p}{p_{0}},
\label{t55}
\end{equation}%
\begin{equation}
s_{T}(x)=\frac{1}{2}\int H(p_{0})\left( m+\frac{p_{T}^{2}}{2\left(
p_{0}+m\right) }\right) \delta \left( \frac{p_{0}+p_{1}}{M}-x\right) \frac{%
d^{3}p}{p_{0}}.  \label{t56}
\end{equation}%
Now, using a standard notation%
\begin{equation}
g_{T}(x)=g_{1}(x)+g_{2}(x),  \label{ta56}
\end{equation}%
we get from the relations (\ref{ta14}) and (\ref{t56}) the equivalence%
\begin{equation}
g_{T}(x)=s_{T}(x).  \label{tb56}
\end{equation}

Comparison of the relations (\ref{t30}) and (\ref{t54}) implies, that%
\begin{equation}
j(x)=\frac{1}{2}V_{0}(x).  \label{tc56}
\end{equation}%
In the Appendix \ref{app4} we have shown, that the distribution $V_{0}(y)$
can be extracted from the relations (\ref{T32}), (\ref{T33}), so we get:%
\begin{equation}
j(x)=\frac{x^{2}+x_{0}^{2}}{2x^{2}}g_{1}(x)+\allowbreak \frac{%
3x^{2}+2xx_{0}+3x_{0}^{2}}{2\left( x+x_{0}\right) ^{2}}\int_{x}^{1}\frac{%
y^{2}-x_{0}^{2}}{y^{3}}g_{1}(y)dy+\int_{x}^{1}\ln \left( \frac{x\left(
y+x_{0}\right) ^{2}}{y\left( x+x_{0}\right) ^{2}}\right) \frac{%
y^{2}-x_{0}^{2}}{y^{3}}g_{1}(y)dy,  \label{T57}
\end{equation}%
\begin{equation}
j(x)=-\allowbreak \frac{x^{2}+x_{0}^{2}}{2x\left( x-x_{0}\right) }g_{2}(x)-%
\frac{1}{2\left( x^{2}-x_{0}^{2}\right) }\int_{x}^{1}\frac{%
6y^{2}x-2yx^{2}+x^{2}x_{0}+2yx_{0}^{2}-x_{0}^{3}}{y^{2}}g_{2}(y)dy.
\label{T58}
\end{equation}%
Then for $x_{0}\rightarrow 0$ we have:%
\begin{equation}
j(x)=\frac{1}{2}g_{1}(x)+\allowbreak \frac{1}{2}\int_{x}^{1}\left( 3+2\ln 
\frac{y}{x}\right) \frac{g_{1}(y)}{y}dy,  \label{t59}
\end{equation}%
\begin{equation}
j(x)=-\frac{1}{2}\allowbreak g_{2}(x)-\frac{1}{x}\int_{x}^{1}\left(
3y-x\right) \frac{g_{2}(y)}{y}dy.  \label{t60}
\end{equation}

Further, comparison of the relations (\ref{t55}) and (\ref{t13}) implies%
\begin{equation}
s_{L}(x)=g_{1}(x)-\frac{1}{2}\int H(p_{0})p_{1}\delta \left( \frac{%
p_{0}+p_{1}}{M}-x\right) \frac{d^{3}p}{p_{0}}.  \label{t61}
\end{equation}%
The last integral can be expressed by means of the function $V_{0}$ (see
Appendix \ref{app5}):%
\begin{equation}
s_{L}(x)=\frac{1}{2}\frac{x^{2}+x_{0}^{2}}{x^{2}}g_{1}(x)-\allowbreak \frac{%
x-3x_{0}}{2\left( x+x_{0}\right) }\int_{x}^{1}\frac{y^{2}-x_{0}^{2}}{y^{3}}%
g_{1}(y)dy+\int_{x}^{1}\ln \left( \frac{x\left( y+x_{0}\right) ^{2}}{y\left(
x+x_{0}\right) ^{2}}\right) \frac{y^{2}-x_{0}^{2}}{y^{3}}g_{1}(y)dy.
\label{T62}
\end{equation}%
A similar procedure gives also%
\begin{equation}
s_{L}(x)=-\frac{1}{2}\frac{x^{2}+x_{0}^{2}}{x\left( x-x_{0}\right) }%
g_{2}(x)+\int_{x}^{1}\left( \frac{2y-x_{0}}{2y^{2}}-\frac{x+2x_{0}}{%
x^{2}-x_{0}^{2}}\right) g_{2}(y)dy  \label{T63}
\end{equation}%
One can easily check, that for $x_{0}\rightarrow 0$ it follows 
\begin{equation}
s_{L}(x)=\frac{1}{2}g_{1}(x)+\int_{x}^{1}\left( \ln \frac{y}{x}-\allowbreak 
\frac{1}{2}\right) \frac{g_{1}(y)}{y}dy,  \label{t64}
\end{equation}%
\begin{equation}
s_{L}(x)=-\frac{1}{2}g_{2}(x)+\int_{x}^{1}\left( \frac{1}{y}-\frac{1}{x}%
\right) g_{2}(y)dy.  \label{t65}
\end{equation}

Finally, combining relations (\ref{t47}),(\ref{T48}) and (\ref{tb56}) one
gets%
\begin{equation}
g_{T}(x)=\frac{x_{0}}{x}g_{1}(x)+\frac{x\left( x+2x_{0}\right) }{\left(
x+x_{0}\right) ^{2}}\int_{x}^{1}\frac{y^{2}-x_{0}^{2}}{y^{3}}g_{1}(y)dy,
\label{t66}
\end{equation}%
\begin{equation}
g_{T}(x)=-\frac{x_{0}}{x-x_{0}}g_{2}(x)-\frac{x+2x_{0}}{x^{2}-x_{0}^{2}}%
\int_{x}^{1}g_{2}(y)dy.  \label{t67}
\end{equation}%
and for $x_{0}\rightarrow 0$ it follows%
\begin{equation}
g_{T}(x)=\int_{x}^{1}\frac{g_{1}(y)}{y}dy=-\frac{1}{x}\int_{x}^{1}g_{2}(y)dy.
\label{t68}
\end{equation}%
In the end of this section let us point out, that the simple relations
above, which define mutual transformations among the functions $%
g_{1}(x),g_{2}(x),j(x),s_{L}(x)$ and $s_{T}(x)$ are obtained on the
assumption that the fermions have some {\it fixed} effective mass $x_{0}$.
In a more general case, for example when the structure functions are related
to the system of fermions with some effective mass spectrum \cite{zav3},
such simple transformations do not exist.

\section{Valence quarks}

\label{sec4}Now let us try to apply the suggested approach to the
description of the proton spin structure. For simplicity, as in \cite{zav4},
we assume:

1) Spin contribution from the sea of quark-antiquark pairs and gluons can be
neglected. Then the three fermions in our approach correspond to the three
proton valence quarks. So, in this scenario, the proton spin is generated
only by the valence quarks. Let us remark, in the cited paper we have shown
that this assumption does not contradict the experimental value $\Gamma _{1}$%
.

2) In accordance with the non-relativistic {\it SU(6)} approach, the spin
contribution of individual valence terms is given as%
\begin{equation}
s_{u}=4/3,\qquad s_{d}=-1/3.  \label{t69}
\end{equation}%
Let us denote momenta distributions of the valence quarks in the target rest
frame by symbols $h_{u}$ and $h_{d}$ with the normalization%
\begin{equation}
\frac{1}{2}\int h_{u}(p_{0})d^{3}p=\int h_{d}(p_{0})d^{3}p=1,  \label{t70}
\end{equation}%
then the generic distribution (\ref{t4}) reads%
\begin{equation}
H(p_{0})=\sum e_{j}^{2}\Delta h_{j}(p_{0})=\left( \frac{2}{3}\right) ^{2}%
\frac{2}{3}h_{u}(p_{0})-\left( \frac{1}{3}\right) ^{2}\frac{1}{3}%
h_{d}(p_{0}).  \label{t71}
\end{equation}

In the papers \cite{zav1}, \cite{zav3}, using a similar approach, we have
studied also the unpolarized structure functions. In particular we have
suggested, that the structure function $F_{2}$, can be in the limit (\ref{t9}%
), expressed as%
\begin{equation}
F_{2}(x)=x^{2}\int K(p_{0})\frac{M}{p_{0}}\delta \left( \frac{p_{0}+p_{1}}{M}%
-x\right) d^{3}p;\qquad K(p_{0})=\sum_{q}e_{q}^{2}h_{q}(p_{0}),  \label{t72}
\end{equation}%
where $h_{q}$ are distributions of the quarks with charges $e_{q}$. For the
valence quarks one can write%
\begin{equation}
F_{2}(x)=\frac{4}{9}xu_{V}(x)+\frac{1}{9}xd_{V}(x),  \label{t73}
\end{equation}%
then the Eq. (\ref{t72}) can be split:%
\begin{equation}
q_{V}(x)=x\int h_{q}(p_{0})\frac{M}{p_{0}}\delta \left( \frac{p_{0}+p_{1}}{M}%
-x\right) d^{3}p;\qquad q=u,d.  \label{t74}
\end{equation}%
In an accordance with the definition (\ref{t30}), in which $h_{q}$ is
inserted instead of $H$, one can write

\begin{equation}
q_{V}(x)=xV_{-1}^{q}(x),  \label{t75}
\end{equation}%
then the relation (\ref{T31}) implies%
\begin{equation}
V_{0}^{q}(x)=\allowbreak \frac{1}{2}\left( x+\frac{x_{0}^{2}}{x}\right)
V_{-1}^{q}(x)+\frac{1}{2}\int_{x}^{1}\left( 1-\frac{x_{0}^{2}}{y^{2}}\right)
V_{-1}^{q}(x)dy,  \label{t76}
\end{equation}%
which after inserting from Eq. (\ref{t75}) gives%
\begin{equation}
V_{0}^{q}(x)=\frac{1}{2}\left[ \allowbreak \left( 1+\frac{x_{0}^{2}}{x^{2}}%
\right) q_{V}(x)+\int_{x}^{1}\left( 1-\frac{x_{0}^{2}}{y^{2}}\right) \frac{%
q_{V}(y)}{y}dy\right] .  \label{t77}
\end{equation}%
Obviously, the function $V_{0}(x)$ generated by distribution (\ref{t71})
according to the definition (\ref{t30}) can be decomposed%
\begin{equation}
V_{0}(x)=\frac{8}{27}V_{0}^{u}(x)-\frac{1}{27}V_{0}^{d}(x)  \label{ta77}
\end{equation}%
and if we define%
\begin{equation}
w_{g}(x)=\frac{8}{27}u_{V}(x)-\frac{1}{27}d_{V}(x),  \label{tb77}
\end{equation}%
then one can check, that inserting $V_{0}$ from the relation (\ref{ta77}) to
the relations (\ref{T32}), (\ref{T33}) with the use of Eqs. (\ref{t77}), (%
\ref{tb77}) gives:%
\begin{equation}
g_{1}(x)=\frac{1}{2}\left[ \allowbreak w_{g}(x)-2\left( x+x_{0}\right)
^{2}\int_{x}^{1}\frac{y-x_{0}}{\left( y+x_{0}\right) ^{3}}\frac{w_{g}(y)}{y}%
dy\right] ,  \label{t78}
\end{equation}%
\begin{equation}
g_{2}(x)=\frac{1}{2}\left[ -\left( 1-\frac{x_{0}}{x}\right) \allowbreak
w_{g}(x)+3\left( x+x_{0}\right) ^{2}\int_{x}^{1}\frac{y-x_{0}}{\left(
y+x_{0}\right) ^{3}}\frac{w_{g}(y)}{y}dy\right] .  \label{t79}
\end{equation}%
Obviously, the structure functions can be split into the two parts,
corresponding to $u$ and $d$ quarks%
\begin{equation}
g_{j}(x)=\left( \frac{2}{3}\right) ^{2}\frac{2}{3}g_{j}^{u}(x)-\left( \frac{1%
}{3}\right) ^{2}\frac{1}{3}g_{j}^{d}(x);\qquad j=1,2,  \label{ta79}
\end{equation}%
where the partial structure functions read:%
\begin{equation}
g_{1}^{q}(x)=\frac{1}{2}\left[ \allowbreak q_{V}(x)-2\left( x+x_{0}\right)
^{2}\int_{x}^{1}\frac{y-x_{0}}{\left( y+x_{0}\right) ^{3}}\frac{q_{V}(y)}{y}%
dy\right] ,  \label{tb79}
\end{equation}%
\begin{equation}
g_{2}^{q}(x)=\frac{1}{2}\left[ -\left( 1-\frac{x_{0}}{x}\right) \allowbreak
q_{V}(x)+3\left( x+x_{0}\right) ^{2}\int_{x}^{1}\frac{y-x_{0}}{\left(
y+x_{0}\right) ^{3}}\frac{q_{V}(y)}{y}dy\right] ;\qquad q=u,d.  \label{tc79}
\end{equation}%
Now we can express the corresponding contributions of different quarks to
the spin distribution functions. Apparently

\begin{equation}
s_{T}^{q}(x)=g_{1}^{q}(x)+g_{2}^{q}(x)=\frac{1}{2}\left[ \frac{x_{0}}{x}%
\allowbreak \allowbreak q_{V}(x)+\left( x+x_{0}\right) ^{2}\int_{x}^{1}\frac{%
y-x_{0}}{\left( y+x_{0}\right) ^{3}}\frac{q_{V}(y)}{y}dy\right] .
\label{t80}
\end{equation}%
Further, after inserting from the relations (\ref{tb79}), (\ref{t77}) and (%
\ref{t75}) to the Eq. (\ref{a38}) one easily gets%
\begin{equation}
s_{L}^{q}(x)=\frac{1}{4}\left[ \allowbreak \left( 1+\frac{x_{0}^{2}}{x^{2}}%
\right) q_{V}(x)-4\left( x+x_{0}\right) ^{2}\int_{x}^{1}\frac{y-x_{0}}{%
\left( y+x_{0}\right) ^{3}}\frac{q_{V}(y)}{y}dy+\int_{x}^{1}\left( 1-\frac{%
x_{0}^{2}}{y^{2}}\right) \frac{q_{V}(y)}{y}dy\right] .  \label{t81}
\end{equation}%
Let us remark, the Eq. (\ref{a38}) is obtained from the generic distribution 
$H(p_{0})$, in its place we now have the distribution $h_{q}(p_{0})$.
Similarly, a comparison of the relations (\ref{tc56}) and (\ref{t77}) gives$%
\allowbreak $%
\begin{equation}
j^{q}(x)=\frac{1}{4}\left[ \allowbreak \left( 1+\frac{x_{0}^{2}}{x^{2}}%
\right) q_{V}(x)+\int_{x}^{1}\left( 1-\frac{x_{0}^{2}}{y^{2}}\right) \frac{%
q_{V}(y)}{y}dy\right] .  \label{t82}
\end{equation}%
Now, the net complete spin distributions can be obtained by adding
individual valence terms with the weights (\ref{t69}), taking into account
their normalization (\ref{t70}). If we define%
\begin{equation}
w_{s}(x)=\frac{2}{3}u_{V}(x)-\frac{1}{3}d_{V}(x),  \label{ta82}
\end{equation}%
then the complete spin distributions can be obtained from the relations (\ref%
{t80}) - (\ref{t82}), in which the distribution $q_{V}$ is replaced by $%
w_{s} $. Then for $x_{0}\rightarrow 0$ we obtain:%
\begin{equation}
g_{1}(x)=\frac{1}{2}\left[ \allowbreak w_{g}(x)-2x^{2}\int_{x}^{1}\frac{%
w_{g}(y)}{y^{3}}dy\right] ,  \label{t83}
\end{equation}%
\begin{equation}
g_{2}(x)=\frac{1}{2}\left[ -\allowbreak \allowbreak
w_{g}(x)+3x^{2}\int_{x}^{1}\frac{w_{g}(y)}{y^{3}}dy\right] ,  \label{t84}
\end{equation}%
\begin{equation}
s_{T}(x)=\frac{1}{2}x^{2}\int_{x}^{1}\frac{w_{s}(y)}{y^{3}}dy,  \label{t85}
\end{equation}%
\begin{equation}
s_{L}(x)=\frac{1}{4}\left[ w_{s}(x)-4x^{2}\int_{x}^{1}\frac{w_{s}(y)}{y^{3}}%
dy+\int_{x}^{1}\frac{w_{s}(y)}{y}dy\right] ,  \label{t86}
\end{equation}%
\begin{equation}
j(x)=\frac{1}{4}\left[ \allowbreak w_{s}(x)+\int_{x}^{1}\frac{w_{s}(y)}{y}dy%
\right] .  \label{t87}
\end{equation}%
Let us note, if one assumes $u_{V}(x)\approx 2d_{V}(x)$, then the following
substitution can be used:%
\begin{equation}
w_{g}(x)\approx \frac{5}{9}\frac{F_{2val}}{x},\qquad w_{s}(x)\approx \frac{%
F_{2val}}{x}.  \label{ta87}
\end{equation}

Further, let us make a remark to the normalization of the distributions
above. To simplify this consideration, we assume the case $x_{0}\rightarrow
0 $. The relation (\ref{t75}) implies%
\begin{equation}
\int_{0}^{1}q_{V}(x)dx=\int_{0}^{1}xV_{-1}^{q}(x)dx.  \label{t88}
\end{equation}%
Since the relation (\ref{T31}) implies%
\begin{equation}
V_{-1}^{q}(x)=\frac{2}{x}V_{0}^{q}(x)-2\int_{x}^{1}\frac{V_{0}^{q}(y)}{y^{2}}%
dy,  \label{t89}
\end{equation}%
then using the relation (\ref{ab37}) one gets%
\begin{equation}
\int_{0}^{1}xV_{-1}^{q}(x)dx=\int_{0}^{1}V_{0}^{q}(x)dx.  \label{t90}
\end{equation}%
From the definition%
\begin{equation}
V_{0}^{q}(x)=\int h_{q}(p_{0})\delta \left( \frac{p_{0}+p_{1}}{M}-x\right)
d^{3}p  \label{t91}
\end{equation}%
one obtains%
\begin{equation}
\int_{0}^{1}V_{0}^{q}(x)dx=\int h_{q}(p_{0})d^{3}p.  \label{t92}
\end{equation}%
This relation combined with (\ref{t90}) and (\ref{t88}) gives%
\begin{equation}
\int_{0}^{1}q_{V}(x)dx=\int h_{q}(p_{0})d^{3}p,  \label{t93}
\end{equation}%
which in an accordance with the normalization (\ref{t70}) implies%
\begin{equation}
\frac{1}{2}\int_{0}^{1}u_{V}(x)dx=\int_{0}^{1}d_{V}(x)dx=1.  \label{t94}
\end{equation}%
Now, one can also check the normalization of the functions (\ref{t83}) - (%
\ref{t87}). Taking into account, that%
\begin{equation}
\int_{0}^{1}w_{s}(x)dx=1,\qquad \int_{0}^{1}w_{g}(x)dx=\frac{5}{9},
\label{t95}
\end{equation}%
then after integration with the use of relation (\ref{ab37}) one gets%
\begin{equation}
\int_{0}^{1}j(x)dx=\frac{1}{2},\qquad
\int_{0}^{1}s_{T}(x)dx=\int_{0}^{1}s_{L}(x)dx=\frac{1}{6},  \label{t96}
\end{equation}%
\begin{equation}
\int_{0}^{1}g_{1}(x)dx=\frac{5}{54},\qquad \int_{0}^{1}g_{2}(x)dx=0.
\label{t97}
\end{equation}%
The meaning of these integrals was discussed in Sec. \ref{sec2}. The first
integral (\ref{ab37}) represents the sum rule on the total proton angular
momentum $J=l+s=1/2$, which has been discussed e.g. in \cite{ter}. The
integrals on the net spin contributions $s_{T},s_{L}$ are correlated with
the $\Gamma _{1}$, which reaches its minimal value for $x_{0}\rightarrow 0$,
as we have discussed in \cite{zav4}, see also Eq. (\ref{t17}) above.

\begin{figure}
\begin{center}
\epsfig{file=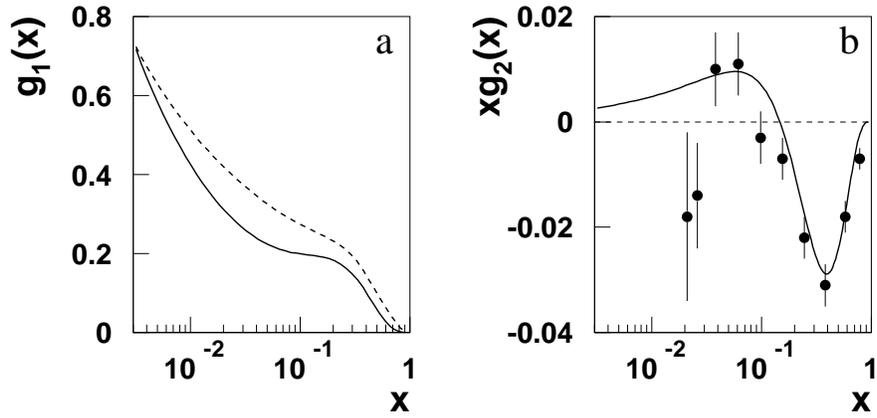, height=7cm}
\end{center}
\caption{Proton spin structure functions. Our calculation, which is
represented by the full lines, is compared with the experimental data:
dashed line ($g_{1}$) and full circles ($g_{2}$).}
\label{gps1}
\end{figure}

\begin{figure}
\begin{center}
\epsfig{file=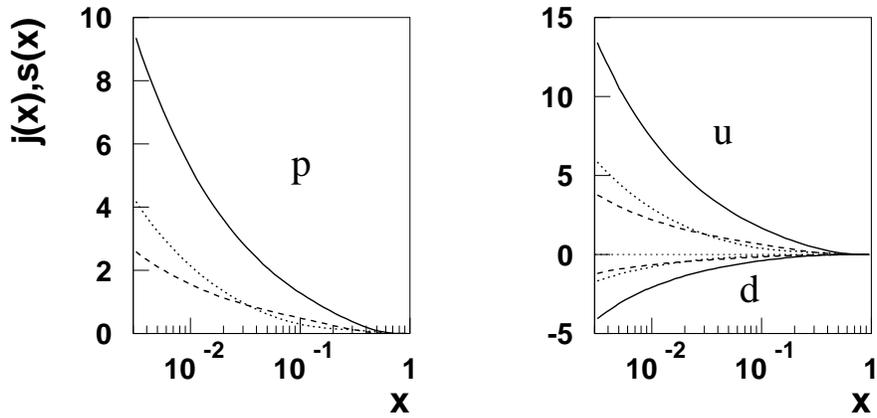, height=7cm}
\end{center}
\caption{Calculation of the spin densities of the valence quarks inside the
proton {\it (left)} and separately the contributions from $u$ and $d$ quarks 
{\it (right)}. Full lines represent the total angular momenta $j$, dotted
and dashed lines correspond to longitudinal and transversal densities $s_{L}$
and $s_{T}$. }
\label{gps2}
\end{figure}

So, the obtained formulas enable us to calculate the spin functions from the
input, in which only the valence distributions are used. For simplicity we
shall now consider only massless quarks ($x_{0}\rightarrow 0$) and for the
valence functions $xu_{V}(x)$ and $xd_{V}(x)$ we use the parameterization
obtained (for $Q^{2}=4GeV^{2}/c^{2}$) by the standard global analysis in %
\cite{msr}. In the Fig. \ref{gps1}{\it a} we have shown the result of our
calculation for $g_{1}$ according to Eq. (\ref{t83}) together with the
experimental data represented by the new parameterization of the world data
on $g_{1}$ \cite{e155g1} for $Q^{2}=4GeV^{2}/c^{2}$. The calculation
qualitatively agrees well with the data, however it is apparent, that the
data are above our curve. This can be connected first of all with our
simplification for $x_{0}\rightarrow 0$, when the $\Gamma _{1}$ is minimal.
In accordance with Eq. (\ref{t97}) we obtain $\Gamma _{1}\doteq 0.093$, but
experimentally $\Gamma _{1}^{p}=0.118\pm 0.004(stat.)\pm 0.007(syst.)$ at $%
Q^{2}=5GeV^{2}/c^{2}$ \cite{e155g1}. Just this difference is exposed in the
figure. Further, in the Fig. \ref{gps1}{\it b} we have shown the $g_{2}$
according to Eq. (\ref{t84}) and the precision measurement recently
published by the E155 Collaboration \cite{e155g2}. The agreement with the
data is very good and it can suggest, that dependence of the function $g_{2}$
on the mass terms is rather weak. At least in our approach $\Gamma _{1}$
does depend on mass, but $\Gamma _{2}=0$ regardless of the mass. In the Fig. %
\ref{gps2} the corresponding spin distributions $j,s_{T},s_{L}$ are shown
for whole proton and also separately for $u$ and $d$ valence quarks
corresponding to the assumed {\it SU(6) }symmetry, which gives the fractions
(\ref{t69}). Figure \ref{gps1}{\it a} and the left part of Fig. \ref{gps2}
also demonstrate, that the $g_{1}(x)$ and $s_{L}(x)$ are not equivalent.
Slightly different shape of the distributions on $s_{T},s_{L}$ is due to
variable $x$, in which longitudinal and transversal (in respect to the beam)
quark momentum components are not involved in a symmetric way. Otherwise,
for given direction of the proton polarization, quark spin density cannot
depend on the direction, in which the probing beam is coming.

In the end, let us remark, that another possible effect, which can in our
approach contribute to an underestimation of $\Gamma _{1}$ is connected with
the assumption\ (\ref{t69}). For example, if one assumes full spin alignment
of the $u$ valence quarks, then%
\begin{equation}
s_{u}=2,\qquad s_{d}=-1  \label{t98}
\end{equation}%
and instead of the generic distribution (\ref{tb77}) one gets%
\begin{equation}
w_{g}(x)=\frac{4}{9}u_{V}(x)-\frac{1}{9}d_{V}(x),\qquad
\int_{0}^{1}w_{g}(x)dx=\frac{7}{9},  \label{t99}
\end{equation}%
which implies $\Gamma _{1}=7/54\doteq 0.13$. Obviously, assuming isotopic
symmetry, the same procedure can also be used for the neutron. For the
isotopic counterparts of the compositions (\ref{t69}) and (\ref{t98}) one
gets $\Gamma _{1}^{n}=0$ and $\Gamma _{1}^{n}=-1/27\doteq -0.037$,
respectively. The composition (\ref{t98}) gives the maximum value $\Gamma
_{1}$ for proton and minimum for neutron.

\section{Summary and conclusion}

\label{sec5}With the use of a consistently covariant version of the naive
QPM we have studied the spin structure functions together with the spin
density distributions for the system of quasifree fermions having fixed
effective mass $x_{0}=m/M$ and the total spin $J=1/2$. The main results can
be summarized as follows.

(1) We have shown that the corresponding spin structure functions $g_{1}(x)$
and $g_{2}(x)$ are mutually connected by a simple transformation. At the
limit $x_{0}\rightarrow 0$ this transformation is identical to the Wanzura -
Wilczek relation for the twist-2 term of the $g_{2}(x)$ approximation. At
the same time for $x_{0}\rightarrow 0$ the relations for the $n-th$ momenta
of the structure functions have been obtained. Their particular cases are
identical to the known sum rules: Wanzura - Wilczek ($n=2,4,6...$), Efremov
- Leader - Teryaev ($n=1$) and Burkhardt - Cottingham ($n=0$). Further, we
have shown how the structure functions are connected with the net spin
densities $s_{L}(x),s_{T}(x)$ and with the density of the total angular
momentum $j(x)$.

(2) The proposed approach has been applied to the description of the proton
spin structure with the assumption that the proton spin is generated only by
the spins and orbital momenta of the valence quarks. Apart from that we have
assumed that the spin contributions from $u$ and $d$ valence quarks can be
defined by the {\it SU(6)} symmetry and for quark effective mass we used
approximation $x_{0}\rightarrow 0$. We have suggested, how one can in this
approach obtain the spin functions from the valence quark distributions.
Then as an input we have used parameterization of the valence terms
resulting from the standard global analysis. On this basis, without any
other free parameter, the proton spin structure functions and related spin
densities have been calculated. Comparison of the obtained structure
functions $g_{1}(x)$ and $g_{2}(x)$ with the experimental measurement
demonstrates that the suggested approach well reproduces the basic features
of the data on the proton spin structure.

To conclude, the results presented in this paper and discussion in \cite%
{zav4} suggest, that both the proton structure functions $g_{1}$ and $g_{2}$
have a simple and natural interpretation even in terms of a naive QPM,
provided that the model is based on a consistently covariant formulation,
which takes into account spheric symmetry connected with the constraint $%
J=1/2$. This is not satisfied for the standard formulation of QPM, which is
based on the one-dimensional kinematics related only to the preferred
reference system (infinite momentum frame). As a result, there is e.g. the
known fact, that the function $g_{2}(x)$ has no well-defined meaning in the
standard naive QPM. In this case it is just a result of the simplified
kinematics and not because of an absence of dynamics.

\bigskip

{\bf Acknowledgements }{\it I would like to thank Anatoli Efremov and Oleg
Teryaev for many useful discussions and valuable comments.}

\appendix

\section{Proof of the relation ({\ref{T31}})}

\label{app1}The integral%
\begin{equation}
F(x)=\int K(p_{0})\delta \left( \frac{p_{0}+p_{1}}{M}-x\right) d^{3}p;\qquad
p_{0}=\sqrt{m^{2}+{\bf p}^{2}}  \label{a1}
\end{equation}%
after the substitution%
\begin{equation}
p_{2}=\sqrt{p_{0}^{2}-p_{1}^{2}-m^{2}}\sin \varphi ,\qquad p_{3}=\sqrt{%
p_{0}^{2}-p_{1}^{2}-m^{2}}\cos \varphi  \label{a2}
\end{equation}%
reads:%
\begin{equation}
F(x)=2\pi \int_{m}^{E_{\max }}K(p_{0})p_{0}\left[ \int_{-H}^{+H}\delta
\left( \frac{p_{0}+p_{1}}{M}-x\right) dp_{1}\right] dp_{0};\qquad H=\sqrt{%
p_{0}^{2}-m^{2}}.  \label{a3}
\end{equation}%
For given $x$ and $p_{0}$ the inner integral contributes only for%
\begin{equation}
p_{1}=Mx-p_{0}  \label{a4}
\end{equation}%
in the limits%
\begin{equation}
-\sqrt{p_{0}^{2}-m^{2}}\leq p_{1}\leq \sqrt{p_{0}^{2}-m^{2}}.  \label{a5}
\end{equation}%
One can check, that the last two conditions are compatible only for%
\begin{equation}
p_{0}\geq \xi \equiv \frac{Mx}{2}+\frac{m^{2}}{2Mx}.  \label{a6}
\end{equation}%
It follows, that Eq. (\ref{a3}) can be simplified%
\begin{equation}
F(x)=2\pi M\int_{\xi }^{E_{\max }}K(p_{0})p_{0}dp_{0}.  \label{a7}
\end{equation}%
According to the relation (\ref{a6}) the parameter $\xi $ is a function of $%
x $ with a minimum at $x_{0}=m/M$, so for the fixed $\xi $ there are the two
roots of $x$,%
\begin{equation}
x_{\pm }=\frac{\xi \pm \sqrt{\xi ^{2}-m^{2}}}{M},  \label{a8}
\end{equation}%
so the Eq. (\ref{a7}) can be rewritten:%
\begin{equation}
F\left( \frac{\xi \pm \sqrt{\xi ^{2}-m^{2}}}{M}\right) =2\pi M\int_{\xi
}^{E_{\max }}K(p_{0})p_{0}dp_{0}.  \label{a9}
\end{equation}%
Then differentiation in respect to $\xi $ gives%
\begin{equation}
F^{\prime }\left( \frac{\xi \pm \sqrt{\xi ^{2}-m^{2}}}{M}\right) \left( 
\frac{1}{M}\pm \frac{\xi }{M\sqrt{\xi ^{2}-m^{2}}}\right) =-2\pi MK(\xi )\xi
\label{a10}
\end{equation}%
and this relation can be, with the use of Eq. (\ref{a8}), applied to the
functions (\ref{t30}):%
\begin{equation}
V_{n}^{\prime }(x_{\pm })x_{\pm }=\mp 2\pi M^{2}H(\xi )\xi \sqrt{\xi
^{2}-m^{2}}\left( \frac{\xi }{M}\right) ^{n},  \label{a11}
\end{equation}%
which with the use of the relation (\ref{a6}) implies the relation (\ref{T31}%
).

\section{ Proof of the relations (\ref{T32}),(\ref{T33})}

\label{app2}In the relations (\ref{t13}) - (\ref{ta14}) one can, due to the $%
\delta -$ function, make the following substitutions. First,%
\begin{equation}
p_{1}=Mx-p_{0}  \label{a12}
\end{equation}%
and then from the relation%
\begin{equation}
p_{T}^{2}=p_{0}^{2}-p_{1}^{2}-m^{2}  \label{a13}
\end{equation}%
one gets%
\begin{equation}
p_{T}^{2}=2Mxp_{0}-\left( M^{2}x^{2}+m^{2}\right) .  \label{a14}
\end{equation}%
Now the relation (\ref{ta14}) can be rewritten:%
\begin{equation}
g_{1}(x)+g_{2}(x)=\frac{1}{2}\left[ x_{0}V_{-1}(x)+\int H(p_{0})\left( \frac{%
Mxp_{0}-\left( M^{2}x^{2}+m^{2}\right) /2}{p_{0}\left( p_{0}+m\right) }%
\right) \delta \left( \frac{p_{0}+p_{1}}{M}-x\right) d^{3}p\right] .
\label{a15}
\end{equation}%
In the next step we expand the fraction%
\begin{equation}
\frac{1}{p_{0}+m}=\frac{1}{p_{0}}\sum_{j=0}^{\infty }\left( -\frac{m}{p_{0}}%
\right) ^{j},  \label{a16}
\end{equation}%
then the relation (\ref{a15}) can be expressed in terms of the functions (%
\ref{t30})%
\begin{equation}
g_{1}(x)+g_{2}(x)=\frac{1}{2}\left[ x_{0}V_{-1}(x)-\frac{x}{x_{0}}%
\sum_{j=1}^{\infty }\left( -x_{0}\right) ^{j}V_{-j}(x)-\frac{x^{2}+x_{0}^{2}%
}{2x_{0}^{2}}\sum_{j=2}^{\infty }\left( -x_{0}\right) ^{j}V_{-j}(x)\right]
\label{a17}
\end{equation}%
\[
=\frac{1}{2}\left[ \left( x+x_{0}\right) V_{-1}(x)-\frac{\left(
x+x_{0}\right) ^{2}}{2x_{0}^{2}}\sum_{j=2}^{\infty }\left( -x_{0}\right)
^{j}V_{-j}(x)\right] . 
\]%
Further, from the relation (\ref{T31}) one obtains%
\begin{equation}
V_{-j}(x)=-\int_{x}^{1}\left( \frac{2y}{y^{2}+x_{0}^{2}}\right)
^{j}V_{0}^{\prime }(y)dy  \label{a18}
\end{equation}%
and because%
\begin{equation}
\sum_{j=2}^{\infty }\left( \frac{-2x_{0}y}{y^{2}+x_{0}^{2}}\right) ^{j}=%
\frac{4x_{0}^{2}y^{2}}{\left( y+x_{0}\right) ^{2}\left(
y^{2}+x_{0}^{2}\right) },  \label{a19}
\end{equation}%
the relation (\ref{a17}) can be modified%
\begin{equation}
g_{1}(x)+g_{2}(x)=\frac{1}{2}\left[ -\left( x+x_{0}\right) \int_{x}^{1}\frac{%
2y}{y^{2}+x_{0}^{2}}V_{0}^{\prime }(y)dy+\frac{\left( x+x_{0}\right) ^{2}}{%
2x_{0}^{2}}\int_{x}^{1}\frac{4x_{0}^{2}y^{2}}{\left( y+x_{0}\right)
^{2}\left( y^{2}+x_{0}^{2}\right) }V_{0}^{\prime }(y)dy\right] .  \label{a20}
\end{equation}%
Now, the integration by parts gives%
\begin{equation}
g_{1}(x)+g_{2}(x)=\frac{1}{2}\left[ \left( x+x_{0}\right) \frac{2x}{%
x^{2}+x_{0}^{2}}-\frac{\left( x+x_{0}\right) ^{2}}{2x_{0}^{2}}\frac{%
4x_{0}^{2}x^{2}}{\left( x+x_{0}\right) ^{2}\left( x^{2}+x_{0}^{2}\right) }%
\right] V_{0}(x)  \label{a21}
\end{equation}%
\[
+\frac{1}{2}\left[ \int_{x}^{1}\left( -\left( x+x_{0}\right) \frac{2\left(
y^{2}-x_{0}^{2}\right) }{\left( y^{2}+x_{0}^{2}\right) ^{2}}+\frac{\left(
x+x_{0}\right) ^{2}}{2x_{0}^{2}}\frac{8x_{0}^{2}y\left(
y^{3}-x_{0}^{3}\right) }{\left( y^{2}+x_{0}^{2}\right) ^{2}\left(
y+x_{0}\right) ^{3}}\right) V_{0}(y)dy\right] 
\]%
and one can check, that after some modifications, the corresponding terms
ahead of $V_{0}$ coincide with the functions (\ref{t35}). The relation (\ref%
{T32}) can be proved by the similar procedure, so we suggest only main steps:

\[
g_{2}(x)=-\frac{1}{2}\left[ xV_{-1}(x)-V_{0}(x)+\int H(p_{0})\left( \frac{%
p_{0}^{2}-3Mxp_{0}+\left( 3M^{2}x^{2}+m^{2}\right) /2}{p_{0}\left(
p_{0}+m\right) }\right) \delta \left( \frac{p_{0}+p_{1}}{M}-x\right) d^{3}p%
\right] , 
\]

\[
g_{2}(x)=-\frac{1}{2}\left[ xV_{-1}(x)-V_{0}(x)+\sum_{j=0}^{\infty }\left(
-x_{0}\right) ^{j}V_{-j}(x)+3\frac{x}{x_{0}}\sum_{j=1}^{\infty }\left(
-x_{0}\right) ^{j}V_{-j}(x)+\frac{3x^{2}+x_{0}^{2}}{2x_{0}^{2}}%
\sum_{j=2}^{\infty }\left( -x_{0}\right) ^{j}V_{-j}(x)\right] 
\]%
\[
=-\frac{1}{2}\left[ -\left( 2x+x_{0}\right) V_{-1}(x)+\frac{3\left(
x+x_{0}\right) ^{2}}{2x_{0}^{2}}\sum_{j=2}^{\infty }\left( -x_{0}\right)
^{j}V_{-j}(x)\right] , 
\]%
\[
g_{2}(x)=-\frac{1}{2}\left[ \left( 2x+x_{0}\right) \int_{x}^{1}\frac{2y}{%
y^{2}+x_{0}^{2}}V_{0}^{\prime }(y)dy-\frac{3\left( x+x_{0}\right) ^{2}}{%
2x_{0}^{2}}\int_{x}^{1}\frac{4x_{0}^{2}y^{2}}{\left( y+x_{0}\right)
^{2}\left( y^{2}+x_{0}^{2}\right) }V_{0}^{\prime }(y)dy\right] . 
\]%
Then, the integration by parts gives:%
\[
g_{2}(x)=-\frac{1}{2}\left[ -\left( 2x+x_{0}\right) \frac{2x}{x^{2}+x_{0}^{2}%
}+\frac{3\left( x+x_{0}\right) ^{2}}{2x_{0}^{2}}\frac{4x_{0}^{2}x^{2}}{%
\left( x+x_{0}\right) ^{2}\left( x^{2}+x_{0}^{2}\right) }\right] V_{0}(x) 
\]%
\[
-\frac{1}{2}\left[ \int_{x}^{1}\left( \left( 2x+x_{0}\right) \frac{2\left(
y^{2}-x_{0}^{2}\right) }{\left( y^{2}+x_{0}^{2}\right) ^{2}}-\frac{3\left(
x+x_{0}\right) ^{2}}{2x_{0}^{2}}\frac{8x_{0}^{2}y\left(
y^{3}-x_{0}^{3}\right) }{\left( y^{2}+x_{0}^{2}\right) ^{2}\left(
y+x_{0}\right) ^{3}}\right) V_{0}(y)dy\right] 
\]%
and one can check, that the corresponding terms ahead of $V_{0}$ coincide
with the functions (\ref{t34}).

\section{Proof of the relations (\ref{T45}) - (\ref{T48})}

\label{app3}First, if we define the functions%
\begin{equation}
f_{1}(y)=\frac{y^{2}-x_{0}^{2}}{\left( y+x_{0}\right) ^{4}x_{0}},\qquad
f_{2}(y)=\frac{y^{2}-x_{0}^{2}}{\left( y^{2}+x_{0}^{2}\right) ^{2}x_{0}},
\label{a22}
\end{equation}%
\begin{equation}
f_{3}(x)=\frac{A(x)}{a(x)}=\frac{x_{0}}{x-x_{0}},\qquad f_{4}(x)=A(x)+a(x)=%
\frac{2x^{2}}{x^{2}+x_{0}^{2}},  \label{a23}
\end{equation}%
then taking into account Eqs. (\ref{T32}) - (\ref{t35}), the Eq. (\ref{t44})
can be rewritten 
\begin{equation}
\left[ -\left( x+x_{0}\right) ^{2}\int_{x}^{1}f_{1}(y)V_{0}(y)dy+\left(
x^{2}-x_{0}^{2}\right) \int_{x}^{1}f_{2}(y)V_{0}(y)dy\right] ^{\prime }
\label{a24}
\end{equation}%
\[
+\left[ -f_{3}(x)3\left( x+x_{0}\right)
^{2}\int_{x}^{1}f_{1}(y)V_{0}(y)dy+f_{3}(x)\left(
3x^{2}+2xx_{0}+x_{0}^{2}\right) \int_{x}^{1}f_{2}(y)V_{0}(y)dy\right]
^{\prime } 
\]%
\[
=c_{1}(x)f_{4}(x)V_{0}(x)+c_{1}(x)2\left( x+x_{0}\right)
^{2}\int_{x}^{1}f_{1}(y)V_{0}(y)dy-c_{1}(x)2\left(
x^{2}+xx_{0}+x_{0}^{2}\right) \int_{x}^{1}f_{2}(y)V_{0}(y)dy 
\]%
\[
-c_{2}(x)a(x)V_{0}(x)-c_{2}(x)3\left( x+x_{0}\right)
^{2}\int_{x}^{1}f_{1}(y)V_{0}(y)dy+c_{2}(x)\left(
3x^{2}+2xx_{0}+x_{0}^{2}\right) \int_{x}^{1}f_{2}(y)V_{0}(y)dy. 
\]%
After differentiating of the l.h.s. one gets%
\begin{equation}
-2\left( x+x_{0}\right)
\int_{x}^{1}f_{1}(y)V_{0}(y)dy+2x\int_{x}^{1}f_{2}(y)V_{0}(y)dy  \label{a25}
\end{equation}%
\[
-3x_{0}\frac{x^{2}-2xx_{0}-3x_{0}^{2}}{\left( x-x_{0}\right) ^{2}}%
\int_{x}^{1}f_{1}(y)V_{0}(y)dy+3x_{0}\frac{x^{2}-2xx_{0}-x_{0}^{2}}{\left(
x-x_{0}\right) ^{2}}\int_{x}^{1}f_{2}(y)V_{0}(y)dy 
\]%
\[
+\left[ \left( x+x_{0}\right) ^{2}f_{1}(x)-\left( x^{2}-x_{0}^{2}\right)
f_{2}(x)+f_{3}(x)3\left( x+x_{0}\right) ^{2}f_{1}(x)-f_{3}(x)\left(
3x^{2}+2xx_{0}+x_{0}^{2}\right) f_{2}(x)\right] V_{0}(x) 
\]%
\[
=c_{1}(x)\frac{2x^{2}}{x^{2}+x_{0}^{2}}V_{0}(x)+c_{1}(x)2\left(
x+x_{0}\right) ^{2}\int_{x}^{1}f_{1}(y)V_{0}(y)dy-c_{1}(x)2\left(
x^{2}+xx_{0}+x_{0}^{2}\right) \int_{x}^{1}f_{2}(y)V_{0}(y)dy 
\]%
\[
-c_{2}(x)2x\frac{x-x_{0}}{x^{2}+x_{0}^{2}}V_{0}(x)-c_{2}(x)3\left(
x+x_{0}\right) ^{2}\int_{x}^{1}f_{1}(y)V_{0}(y)dy+c_{2}(x)\left(
3x^{2}+2xx_{0}+x_{0}^{2}\right) \int_{x}^{1}f_{2}(y)V_{0}(y)dy. 
\]%
This equality contains the three linearly independent terms%
\begin{equation}
V_{0}(x),\qquad \int_{x}^{1}f_{1}(y)V_{0}(y)dy,\qquad
\int_{x}^{1}f_{2}(y)V_{0}(y)dy  \label{a26}
\end{equation}%
and comparison of these terms on both the sides gives the equations:%
\begin{equation}
c_{1}(x)x-c_{2}(x)\left( x-x_{0}\right) =-\frac{x+2x_{0}}{x+x_{0}},
\label{a27}
\end{equation}%
\begin{equation}
2c_{1}(x)-3c_{2}(x)=\allowbreak -\frac{2x^{2}-xx_{0}-7x_{0}^{2}}{\left(
x+x_{0}\right) \left( x-x_{0}\right) ^{2}},  \label{a28}
\end{equation}%
\begin{equation}
2c_{1}(x)\left( x^{2}+xx_{0}+x_{0}^{2}\right) -c_{2}(x)\left(
3x^{2}+2xx_{0}+x_{0}^{2}\right) =-\allowbreak \frac{%
2x^{3}-x^{2}x_{0}-4xx_{0}^{2}-3x_{0}^{3}}{\left( x-x_{0}\right) ^{2}}.
\label{a29}
\end{equation}%
One can check, that these three equations with the two unknown $c_{1},c_{2}$
are dependent and solvable, giving the solution (\ref{t46}).

Now, we can solve differential equation (\ref{T45}) for $g_{1}$ or $g_{2}$.
Its homogenous version for $g_{1}$ reads:%
\begin{equation}
g_{1}^{\prime }(x)+\frac{x^{2}+4xx_{0}+x_{0}^{2}}{\left(
x^{2}-x_{0}^{2}\right) \left( x+2x_{0}\right) }g_{1}(x)=0,  \label{a30}
\end{equation}%
which has the solution%
\begin{equation}
g_{1}(x)=C\frac{x+2x_{0}}{x^{2}-x_{0}^{2}}.  \label{a31}
\end{equation}%
Nonhomogeneous term [the part of Eq. (\ref{T45}) involving $g_{2}$] gives
the equation for the function $C(x)$%
\begin{equation}
C^{\prime }(x)\frac{x+2x_{0}}{x^{2}-x_{0}^{2}}=-\frac{x_{0}\left(
x^{2}+xx_{0}+4x_{0}^{2}\right) }{\left( x-x_{0}\right) \left(
x^{2}-x_{0}^{2}\right) \left( x+2x_{0}\right) }g_{2}(x)-\left( \frac{x}{%
x-x_{0}}g_{2}(x)\right) ^{\prime },  \label{a32}
\end{equation}%
which has the solution%
\begin{equation}
C(x)=-x\frac{x+x_{0}}{x+2x_{0}}g_{2}(x)-\int_{x}^{1}g_{2}(y)dy.  \label{a33}
\end{equation}%
After inserting into Eq. (\ref{a31}) one gets the relation (\ref{T48}). The
inverse relation (\ref{t47}) can be proved in a similar way.

\section{Proof of the relations (\ref{T57}), (\ref{T58})}

\label{app4}The relations (\ref{T32}) and (\ref{T33})

\[
g_{2}(x)=-\left[ x\frac{x-x_{0}}{x^{2}+x_{0}^{2}}V_{0}(x)+\int_{x}^{1}\left( 
\frac{3\left( x+x_{0}\right) ^{2}}{\left( y+x_{0}\right) ^{4}}-\frac{%
3x^{2}+2xx_{0}+x_{0}^{2}}{\left( y^{2}+x_{0}^{2}\right) ^{2}}\right) \frac{%
y^{2}-x_{0}^{2}}{2x_{0}}V_{0}(y)dy\right] , 
\]%
\[
g_{1}(x)+g_{2}(x)=\left[ \frac{xx_{0}}{x^{2}+x_{0}^{2}}V_{0}(x)+\int_{x}^{1}%
\left( -\frac{\left( x+x_{0}\right) ^{2}}{\left( y+x_{0}\right) ^{4}}+\frac{%
x^{2}-x_{0}^{2}}{\left( y^{2}+x_{0}^{2}\right) ^{2}}\right) \frac{%
y^{2}-x_{0}^{2}}{2x_{0}}V_{0}(y)dy\right] 
\]%
can be combined in such a way that the second terms in the integrals cancel:%
\begin{equation}
\left( 3x^{2}+2xx_{0}+x_{0}^{2}\right) g_{1}(x)+2\left(
x^{2}+xx_{0}+x_{0}^{2}\right) g_{2}(x)  \label{a34}
\end{equation}%
\[
=\left( x+2x_{0}\right) xV_{0}(x)-\left( x+2x_{0}\right) \left(
x+x_{0}\right) ^{2}\int_{x}^{1}\frac{y^{2}-x_{0}^{2}}{\left( y+x_{0}\right)
^{4}}V_{0}(y)dy. 
\]%
Further, this equation can be modified%
\[
\frac{x}{\left( x+x_{0}\right) ^{2}}V_{0}(x)-\int_{x}^{1}\frac{%
y^{2}-x_{0}^{2}}{\left( y+x_{0}\right) ^{4}}V_{0}(y)dy=\frac{%
3x^{2}+2xx_{0}+x_{0}^{2}}{\left( x+2x_{0}\right) \left( x+x_{0}\right) ^{2}}%
g_{1}(x)+\frac{2\left( x^{2}+xx_{0}+x_{0}^{2}\right) }{\left(
x+2x_{0}\right) \left( x+x_{0}\right) ^{2}}g_{2}(x), 
\]%
which implies a differential equation for $V_{0}(x)$:%
\begin{equation}
\left( \frac{x}{\left( x+x_{0}\right) ^{2}}V_{0}(x)\right) ^{\prime }+\frac{%
x^{2}-x_{0}^{2}}{\left( x+x_{0}\right) ^{4}}V_{0}(x)=\left( \frac{%
3x^{2}+2xx_{0}+x_{0}^{2}}{\left( x+2x_{0}\right) \left( x+x_{0}\right) ^{2}}%
g_{1}(x)+\frac{2\left( x^{2}+xx_{0}+x_{0}^{2}\right) }{\left(
x+2x_{0}\right) \left( x+x_{0}\right) ^{2}}g_{2}(x)\right) ^{\prime }.
\label{a35}
\end{equation}%
The corresponding homogenous equation%
\begin{equation}
\frac{x}{\left( x+x_{0}\right) ^{2}}V_{0}^{\prime }(x)=0  \label{aa35}
\end{equation}%
gives the solution%
\begin{equation}
V_{0}(x)=C  \label{ab35}
\end{equation}%
and for $C(x)$ we have the equation%
\begin{equation}
\frac{x}{\left( x+x_{0}\right) ^{2}}C^{\prime }(x)=\left( \frac{%
3x^{2}+2xx_{0}+x_{0}^{2}}{\left( x+2x_{0}\right) \left( x+x_{0}\right) ^{2}}%
g_{1}(x)+\frac{2\left( x^{2}+xx_{0}+x_{0}^{2}\right) }{\left(
x+2x_{0}\right) \left( x+x_{0}\right) ^{2}}g_{2}(x)\right) ^{\prime }.
\label{a36}
\end{equation}%
The solution reads:%
\begin{equation}
V_{0}(x)=C(x)=\frac{3x^{2}+2xx_{0}+x_{0}^{2}}{\left( x+2x_{0}\right) x}%
g_{1}(x)+\frac{2\left( x^{2}+xx_{0}+x_{0}^{2}\right) }{\left(
x+2x_{0}\right) x}g_{2}(x)  \label{a37}
\end{equation}%
\[
+\int_{x}^{1}\frac{\left( 3y^{2}+2yx_{0}+x_{0}^{2}\right) \left(
y-x_{0}\right) }{\left( y+x_{0}\right) \left( y+2x_{0}\right) y^{2}}%
g_{1}(y)dy+\int_{x}^{1}\frac{2\left( y^{2}+yx_{0}+x_{0}^{2}\right) \left(
y-x_{0}\right) }{\left( y+x_{0}\right) \left( y+2x_{0}\right) y^{2}}%
g_{2}(y)dy. 
\]%
After inserting from Eq. (\ref{t47}) one gets%
\[
V_{0}(x)=\frac{3x^{2}+2xx_{0}+x_{0}^{2}}{\left( x+2x_{0}\right) x}g_{1}(x)+%
\frac{2\left( x^{2}+xx_{0}+x_{0}^{2}\right) }{\left( x+2x_{0}\right) x}%
\left( -\frac{x-x_{0}}{x}g_{1}(x)+\frac{x\left( x+2x_{0}\right) }{\left(
x+x_{0}\right) ^{2}}\int_{x}^{1}\frac{y^{2}-x_{0}^{2}}{y^{3}}%
g_{1}(y)dy\right) 
\]%
\[
+\int_{x}^{1}\frac{\left( 3y^{2}+2yx_{0}+x_{0}^{2}\right) \left(
y-x_{0}\right) }{\left( y+x_{0}\right) \left( y+2x_{0}\right) y^{2}}%
g_{1}(y)dy 
\]%
\[
+\int_{x}^{1}\frac{2\left( y^{2}+yx_{0}+x_{0}^{2}\right) \left(
y-x_{0}\right) }{\left( y+x_{0}\right) \left( y+2x_{0}\right) y^{2}}\left( -%
\frac{y-x_{0}}{y}g_{1}(y)+\frac{y\left( y+2x_{0}\right) }{\left(
y+x_{0}\right) ^{2}}\int_{y}^{1}\frac{z^{2}-x_{0}^{2}}{z^{3}}%
g_{1}(z)dz\right) dy, 
\]%
The double integral is calculated according to the formula%
\begin{equation}
\int_{x}^{1}a(y)\left( \int_{y}^{1}b(z)dz\right) dy=\int_{x}^{1}\left(
A(y)-A(x)\right) b(y)dy;\qquad A^{\prime }(x)=a(x),  \label{ab37}
\end{equation}%
then after collecting the corresponding terms with $g_{1}$ we obtain with
the use of Eq. (\ref{tc56}) the relation (\ref{T57}). A similar procedure
with inserting from Eq. (\ref{T48}) into Eq. (\ref{a37}) gives the relation (%
\ref{T58}).

\section{Proof of the relations (\ref{T62}), (\ref{T63})}

\label{app5}In the relation (\ref{t61}) one can, due to the $\delta $
function, make the substitution $p_{1}=Mx-p_{0}$. Then, using the definition
(\ref{t30}), one obtains 
\begin{equation}
s_{L}(x)=g_{1}(x)+\frac{1}{2}V_{0}(x)-\frac{1}{2}xV_{-1}(x).  \label{a38}
\end{equation}%
Further, the relation (\ref{T31}) implies%
\begin{equation}
V_{-1}(x)=\frac{2x}{x^{2}+x_{0}^{2}}V_{0}(x)-2\int_{x}^{1}\frac{%
y^{2}-x_{0}^{2}}{\left( y^{2}+x_{0}^{2}\right) ^{2}}V_{0}(y)dy,  \label{a39}
\end{equation}%
which after inserting into Eq. (\ref{a38}) gives 
\begin{equation}
s_{L}(x)=g_{1}(x)-\frac{x^{2}-x_{0}^{2}}{2\left( x^{2}+x_{0}^{2}\right) }%
V_{0}(x)+x\int_{x}^{1}\frac{y^{2}-x_{0}^{2}}{\left( y^{2}+x_{0}^{2}\right)
^{2}}V_{0}(y)dy.  \label{a40}
\end{equation}%
After inserting $V_{0}=2j(x)$ and using the relation (\ref{T57}) we obtain%
\[
s_{L}(x)=\allowbreak \frac{1}{2}\frac{x^{2}+x_{0}^{2}}{x^{2}}g_{1}(x)-\frac{%
x^{2}-x_{0}^{2}}{2\left( x^{2}+x_{0}^{2}\right) }\allowbreak \frac{%
3x^{2}+2xx_{0}+3x_{0}^{2}}{\left( x+x_{0}\right) ^{2}}\int_{x}^{1}\frac{%
y^{2}-x_{0}^{2}}{y^{3}}g_{1}(y)dy 
\]%
\[
-\frac{x^{2}-x_{0}^{2}}{\left( x^{2}+x_{0}^{2}\right) }\int_{x}^{1}\ln
\left( \frac{x\left( y+x_{0}\right) ^{2}}{y\left( x+x_{0}\right) ^{2}}%
\right) \frac{y^{2}-x_{0}^{2}}{y^{3}}g_{1}(y)dy+x\int_{x}^{1}\frac{%
y^{2}-x_{0}^{2}}{\left( y^{2}+x_{0}^{2}\right) y^{2}}g_{1}(y)dy 
\]

\[
+x\int_{x}^{1}\frac{y^{2}-x_{0}^{2}}{\left( y^{2}+x_{0}^{2}\right) ^{2}}%
\allowbreak \frac{3y^{2}+2yx_{0}+3x_{0}^{2}}{\left( y+x_{0}\right) ^{2}}%
\int_{y}^{1}\frac{z^{2}-x_{0}^{2}}{z^{3}}g_{1}(z)dzdy 
\]%
\[
+2x\int_{x}^{1}\frac{y^{2}-x_{0}^{2}}{\left( y^{2}+x_{0}^{2}\right) ^{2}}%
\int_{y}^{1}\ln \left( \frac{y\left( z+x_{0}\right) ^{2}}{z\left(
y+x_{0}\right) ^{2}}\right) \frac{z^{2}-x_{0}^{2}}{z^{3}}g_{1}(z)dzdy. 
\]%
Then calculation of the double integrals with the use of relation (\ref{ab37}%
) and collecting the corresponding terms with $g_{1}$ give the relation (\ref%
{T62}).

Further, one can insert $g_{1}$ from the relation (\ref{T48}) into relation (%
\ref{T62}), then a similar procedure with double integrals gives the
relation (\ref{T63}).

\newpage

\end{document}